%% file: ms.tex
\documentclass[final,onefignum,onetabnum]{siamonline190516}

\input{macros.tex}
\newcommand{\localfigwidth}{.6\textwidth}
\newcommand{\dropcaplocal}[1]{#1}

\ifpdf
  \DeclareGraphicsExtensions{.eps,.pdf,.png,.jpg}
\else
  \DeclareGraphicsExtensions{.eps}
\fi

\usepackage{enumitem}
\setlist[enumerate]{leftmargin=.5in}
\setlist[itemize]{leftmargin=.5in}

\newsiamremark{remark}{Remark}
\newsiamremark{hypothesis}{Hypothesis}
\crefname{hypothesis}{Hypothesis}{Hypotheses}
\newsiamthm{claim}{Claim}

\headers{Exploiting Asynchronous Priority Scheduling in Parallel Eikonal Solvers}{I. Henriksen, B. You, K. Pingali}

\title{Exploiting Asynchronous Priority Scheduling in Parallel Eikonal Solvers\thanks{Compiled \today.
\funding{This research was supported by NSF grants 1618425, 1705092, and
1725322, and by the Department of Energy and National Nuclear Security Administration under Award Number DE-NA0003969.}}}

\author{Ian Henriksen\thanks{Oden Institute for Computational Engineering and Sciences, University of Texas at Austin, Austin, TX (\email{ian@oden.utexas.edu}).}
\and Bozhi You\thanks{Department of Computer Science, University of Texas at Austin, Austin, TX(\email{youbozhi@cs.utexas.edu}, \email{pingali@cs.utexas.edu}).}
\and Keshav Pingali\footnotemark[1]\footnotemark[2]}

\usepackage{amsopn}

\ifpdf
\hypersetup{
  pdftitle={Exploiting Asynchronous Priority Scheduling in Parallel Eikonal Solvers},
  pdfauthor={I. Henriksen, B. You, K. Pingali}
}
\fi

\begin{document}

\maketitle

\begin{abstract}
\input{abstract.tex}
\end{abstract}



\section*{Significance Statement}
\input{significance_statement.tex}

\section*{Introduction}
\input{introduction.tex}

\input{background.tex}

\input{solvers.tex}

\input{schedule.tex}

\input{precision.tex}

\input{results.tex}

\input{conclusion.tex}

\section*{Methods}
\input{methods.tex}



\bibliographystyle{siamplain}
\bibliography{fmm}

\end{document}

%% file: macros.tex
\usepackage{soul}
\newcommand{\TODO}[2][]{\ifthenelse{\isempty{#1}}
  {[TODO] \hl{#2}}
  {[TODO] \hl{#2 (#1)}}
}

\usepackage{amsfonts}
\usepackage{amsmath}
\usepackage{algorithm,algpseudocode}
\usepackage{booktabs}
\usepackage{mathtools}
\usepackage{verbatim}
\usepackage{xcolor}

\makeatletter
\newcommand{\ALOOP}[1]{\ALC@it\algorithmicloop\ #1%
  \begin{ALC@loop}}
\newcommand{\ENDALOOP}{\end{ALC@loop}\ALC@it\algorithmicendloop}

\newcommand{\algorithmiccontinue}{\textbf{continue}}

\newcommand{\Continue}{\State \algorithmiccontinue}
\makeatother

\newcommand{\pr}[1]{\left(#1\right)}
\newcommand{\br}[1]{\left[#1\right]}
\newcommand{\set}[1]{\left\{#1\right\}}

\newcommand{\norm}[1]{\left\|#1\right\|}
\newcommand{\abs}[1]{\left|#1\right|}

\newcommand{\floor}[1]{\left\lfloor #1 \right\rfloor}

\newcommand{\obimfmm}{Asynchronous Marching Method}

\def \arrivaltime {\operatorname{t}}
\def \speedfunction {\operatorname{v}}

\newcommand\ofpoint[2][]{\ifthenelse{\isempty{#1}}
{\pr{#2}}
{\pr{{#2}_{#1}}}}
\renewcommand\ofpoint[2][]{_{#1}}

%% file: abstract.tex
Numerical solutions to the Eikonal equation are computed using variants of the fast marching method, the fast sweeping method, and the fast iterative method. In this paper, we provide a unified view of these algorithms that highlights their similarities and suggests a wider class of Eikonal solvers. We then use this framework to justify applying concurrent priority scheduling techniques to Eikonal solvers.
We demonstrate that doing so results in good parallel performance for a problem from seismology. We explain why existing Eikonal solvers may produce different results despite using the same differencing scheme and demonstrate techniques to address these discrepancies.
These techniques allow us to obtain deterministic output from our asynchronous fine-grained parallel Eikonal solver.

%% file: significance_statement.tex
Existing Eikonal solvers utilize parallel hardware by scheduling local updates with various ordering constraints.
Here we show that, provided round-off error is handled carefully, local updates can be applied in any order to compute the desired result.
This allows us to use modern concurrent priority scheduling techniques to obtain excellent parallel performance in an Eikonal solver.
The novel techniques introduced here for handling round-off error show that deterministic results can be obtained from a fundamentally unordered computation even in the presence of floating point arithmetic.
They also provide a guide for obtaining parallel performance for the broad class of problems where the fast marching method{\textemdash}an existing technique for solving Eikonal equations{\textemdash}has already been applied.

%% file: introduction.tex
\dropcaplocal{T}he Eikonal equation is a nonlinear PDE with applications in seismology~~\cite{fmm_second_order}, path planning~\cite{fmm_marine}, computer vision~\cite{fmm_computer_vision}, and many other fields. A first-order upwind finite differencing scheme~\cite{finite_diff} is often used to obtain a numerical solution to the Eikonal equation. The fast marching method~\cite{fmm_original}, fast sweeping method~\cite{fast_sweeping}, and fast iterative method~\cite{fast_iterative} are all based on this differencing scheme. The fast marching method is based on a priority queue that follows a {\em narrow band} of discrete points that "marches" across the domain of interest. It was first introduced by Tsitsiklis in~\cite{fmm_original} and many variants~\cite{fmm_main}~\cite{fmm_second_order}~\cite{fmm_second_order_unstructured} have been developed since.
The fast sweeping method~\cite{fast_sweeping} is another algorithm that sweeps across the domain of interest in alternating directions to achieve a solution. The fast iterative method~\cite{fast_iterative} is a round-based algorithm that shares the idea of a {\em narrow band} but does not rely on a priority queue.

The fast marching method has historically been regarded as an inherently sequential algorithm~\cite{marching_sweeping_multilevel}.
The fast sweeping method and fast iterative method are parallelizable~\cite{fast_sweeping_parallel_1}~\cite{fast_sweeping_parallel_2}~\cite{fast_iterative}. In particular, the fast iterative method has been adapted to run in single-GPU~\cite{fim_meshes}~\cite{fim_meshes_3d} and multi-GPU~\cite{fim_mgpu} settings.


The fast marching, fast sweeping, and fast iterative methods have been classically treated as distinct Eikonal solvers.
In this paper, we argue that to understand and further develop parallelism in Eikonal solvers, we need to study them in a unified framework.
To this end, we draw on the {\em operator formulation} introduced by Pingali et al. in~\cite{the_tao}.
The operator formulation provides an abstraction in which algorithms are viewed as a local update operator and a corresponding parallel execution schedule. It has been used to describe and classify graph algorithms including single source shortest path (SSSP) algorithms and has motivated optimized SSSP implementations like the one introduced in~\cite{obim}. Using this approach, we can describe Eikonal solvers in a unified way and better analyze parallelism strategies available in this problem.

Our unified description relies heavily on the similarities between Eikonal solvers and the SSSP problem.
For example, existing implementations of the fast marching method are sequential because they rely on a centralized priority queue.
Djikstra's algorithm for SSSP similarly relies on a priority queue to schedule local updates.
Though concurrent priority queues solve this problem by maintaining per-thread queues with either lock-based or lock-free synchronization, Lenharth et al. in~\cite{obim} show that {\em priority queues are not good concurrent priority schedulers}.
To solve this problem for SSSP,~\cite{obim} develops the {\em Ordered By Integer Metric} (OBIM) scheduler.
OBIM is not a concurrent priority queue.
It is a {\em concurrent priority scheduler} with more relaxed semantics that enable it to avoid constant contention for earliest priority work items while still using priority information for scheduling.
Here we leverage the similarities between Djikstra's algorithm and the fast marching method to adapt the OBIM scheduler for use in a shared-memory paralel Eikonal solver.

There is already a large body of work on domain-decomposition methods for parallelizing the fast marching method~\cite{fmm_parallel_1}~\cite{fmm_parallel_2}~\cite{fmm_parallel_3}~\cite{fmm_parallel_4}~\cite{fmm_parallel_5}~\cite{fmm_parallel_6}~\cite{fmm_parallel_7}~\cite{fmm_parallel_latest}.
~\cite{fast_sweeping_parallel_1} first applied domain decomposition to the fast sweeping method.
~\cite{fmm_parallel_6} explored using a domain decomposition approach at the shared-memory level to obtain good performance with a CPU.
~\cite{fmm_parallel_7} introduced a distributed priority scheduling technique that allowed a distributed fast marching Eikonal solver to scale well in a distributed setting.
Our fine-grained parallel scheme is largely orthogonal to the existing domain decomposition approaches for the fast marching method.
Combining the two approaches in a distributed setting is an area for future work.

The well-known causal properties~\cite{sethian_fmm_survey} that allow use of the fast marching method are critical in enabling the relaxed priority-aware execution described here.
Prior work~\cite{fmm_sloppy}~\cite{fmm_parallel_survey} has noted, however, that different Eikonal solvers may produce different results even if they are implemented using the same update operator.
Here we show that these differences in output arise due to nonmonotonicity in the update operator when it is computed using finite precision arithmetic.
When updates are scheduled asynchronously, these precision issues can cause nondeterministic errors in the solver's output.
To address this we present techniques for mitigating or eliminating these errors.
The technique used here for performing truly monotonic updates is described in terms of the first-order upwind differencing scheme from~\cite{finite_diff}, however it provides a roadmap for how similar results might be achieved for other more sophisticated numerical methods.

For evaluation, we implemented our algorithm using Galois~\cite{galois}, a state-of-the-art parallel graph processing system, and evaluated its performance using input data from the Elastic Marmousi model~\cite{marmousi2}.
The results show good parallel scaling in the shared memory setting and show that the techniques for producing deterministic output introduce minimal overhead.
The empirical results show significant improvements over existing serial and parallel methods in both running time and parallel scaling.


In this paper, we first give a brief introduction to the Eikonal equation and the finite difference scheme.
We then introduce TAO{\textemdash}the methodology we used to analyze existing Eikonal solvers.
Based on that perspective we then introduce the \obimfmm (\obimfmm[abbr]) algorithm.
After that we discuss the floating point precision issues and the determinism of the solution.
We finally show some empirical results.

%% file: background.tex
\section*{Background}
\label{sec:background}

\subsection*{The Eikonal Equation}
\label{sec:eikonal}

Given an open set $\Omega$ in $\mathbb{R}^2$ with well-behaved boundary $\partial \Omega$ and a real-valued velocity function $\speedfunction > 0$ on $\overline{\Omega}$ (i.e. $\Omega$ and its boundary), we consider the Dirichlet boundary value problem (BVP) for computing the wavefront arrival time $\arrivaltime$ at each point in $\overline{\Omega}$.

\begin{equation}
\label{eq:eikonal_bvp}
\begin{aligned}
\norm{\nabla \arrivaltime\pr{x, y}}_2
	&= \frac{1}{\speedfunction\pr{x, y}}
	& \pr{x, y} \in \Omega \\
\arrivaltime\pr{x, y}
	&= 0
	& \pr{x, y} \in \partial \Omega
\end{aligned}
\end{equation}

\cite{eikonal_existence_uniqueness} gives a treatment of the existence and uniqueness of solutions for this problem.
Though more general treatments of the existence and uniqueness of solutions to (\ref{eq:eikonal_bvp}) are possible, here it is sufficient to note that there is a unique viscosity solution $\arrivaltime \in C\pr{\overline{\Omega}}$ provided that $\speedfunction \in C\pr{\overline{\Omega}}$ ($C\pr{\overline{\Omega}}$ is the set of continuous functions on $\overline{\Omega}$).
 
 \subsection*{Upwind Finite Difference Scheme}
\label{sec:finite_diff}
A well known first-order upwind finite difference scheme \cite{finite_diff} builds a numerical approximation to Eq. \ref{eq:eikonal_bvp} on a regular discretization of $\Omega$.
It approximates the gradient at each point by taking the maximum between $0$ and each of the forward and backward difference approximations to the derivative at that point.

Let $\Omega = X \times Y \subset \mathbb{R}^2$ be a square domain with a corresponding discrete grid of evenly spaced points with spacing $h$ where each point $n$ is associated with some notations listed in Table~\ref{tab:notations}.
\begin{table}[ht!]
\centering
\caption{Notations for each point $n$.}
\begin{tabular}{|ll|}
\hline
$i, j$ & Indices corresponding to a given grid point \\
$x_i, y_j$ & Cartesian coordinates of a grid point \\
$n_{i, j}$ & The point $\pr{x_i, y_j}$ \\
$\arrivaltime_{i, j}$ & Arrival time at $n$, equiv. to $\arrivaltime\pr{x_i, y_j}$ \\
$\speedfunction_{i, j}$ & Speed of propagation at $n$, equiv. to $\speedfunction\pr{x_i, y_j}$ \\
\hline
\end{tabular}
\label{tab:notations}
\end{table}
The 2D form of Eq. \ref{eq:eikonal_bvp} is approximated at each $n_{i,j}$ as
\begin{equation}
\label{eq:finite_difference}
{\left \lVert
	\left( \begin{aligned}
		\max\{D^{-x}\arrivaltime\ofpoint[i,j]{n}, -D^{+x}\arrivaltime\ofpoint[i,j]{n}, 0\} \\
		\max\{D^{-y}\arrivaltime\ofpoint[i,j]{n}, -D^{+y}\arrivaltime\ofpoint[i,j]{n}, 0\}
	\end{aligned} \right)
\right \rVert}_2
= \frac{1}{\speedfunction\ofpoint[i,j]{n}}
\end{equation}
with
\begin{equation}
\label{eq:fin_diff_eg}
\begin{aligned}
D^{-x}\arrivaltime\ofpoint[i,j]{n} & = \frac{\arrivaltime\ofpoint[i,j]{n} - \arrivaltime\ofpoint[i-1,j]{n}}{h}\\
D^{+x}\arrivaltime\ofpoint[i,j]{n} & = \frac{\arrivaltime\ofpoint[i+1,j]{n} - \arrivaltime\ofpoint[i,j]{n}}{h}
\end{aligned}
\end{equation}
and $D^{-y}$ and $D^{+y}$ defined similarly in the $y$ dimension.

At each $i, j$ pair, let ${\arrivaltime_V}\ofpoint[i,j]{n} = \min\{\arrivaltime\ofpoint[i,j-1]{n}, \arrivaltime\ofpoint[i,j+1]{n}\}$, and ${\arrivaltime_H}\ofpoint[i,j]{n} = \min\{\arrivaltime\ofpoint[i-1,j]{n}, \arrivaltime\ofpoint[i+1,j]{n}\}$.
When $\abs{{\arrivaltime_H}\ofpoint[i,j]{n} - {\arrivaltime_V}\ofpoint[i,j]{n}} < \frac{h}{\speedfunction\ofpoint[i,j]{n}}$ neither of the maximum operations yield $0$ so the differencing scheme can be rewritten at non-boundary points as
\begin{equation}
\label{eq:difference_as_root}
\pr{\arrivaltime\ofpoint[i,j]{n} - {\arrivaltime_H}\ofpoint[i,j]{n}}^2 + \pr{\arrivaltime\ofpoint[i,j]{n} - {\arrivaltime_V}\ofpoint[i,j]{n}}^2 = \frac{h^2}{\speedfunction\ofpoint[i,j]{n}}
\end{equation}
In cases where $\abs{{\arrivaltime_H}\ofpoint[i,j]{n} - {\arrivaltime_V}\ofpoint[i,j]{n}} \geq \frac{h}{\speedfunction\ofpoint[i,j]{n}}$, the relation $\arrivaltime\ofpoint[i,j]{n} = \operatorname{min}\pr{{\arrivaltime_H}\ofpoint[i,j]{n}, {\arrivaltime_V}\ofpoint[i,j]{n}}$ yields a solution instead.

This differencing scheme is both causal and monotone \cite{sethian_fmm_survey}
It is monotone in the sense that the value at a given node can at most decrease in response to a decrease in the estimated value at one of its neighbors.
It is causal in the sense that the value at each node ''depends`` only on the values of lower-valued neighbors.
These two deterministic properties will be discussed at length later in this paper.
Both are key when showing that deterministic approximate solutions to the Eikonal equation can be computed by updating values in any order.

The fast marching, fast iterative, and fast sweeping methods all rely on this differencing scheme.
The fast marching method relies on a priority queue to track a band of points that marches across the domain of interest.
The fast iterative method tracks a similar band of points, however instead of using a priority queue, it proceeds in rounds, processing all the nodes in the band in each round.
The fast sweeping method instead iterates over the whole domain repeatedly, ignoring nodes that do not need to be updated.
Since the dependencies between

%% file: solvers.tex
\section*{TAO: A Unified View of Existing Eikonal Solvers}
\label{sec:rel_works}

We use the \textit{The Operator Formulation of Algorithms}, described by Pingali et. al. in \cite{the_tao}, to give a unified description of the three methods for solving the Eikonal equation. This methodology is called \textit{TAO analysis}. This unified description characterizes the design space for known Eikonal solvers and suggests new algorithms such as the approximate priority scheduling technique explored later in this paper.

In the Operator Formulation, an algorithm is decomposed into a global view and a local view of the computation. The local view of the algorithm is described by an \textit{operator}, which is applied at an \textit{active node} in the graph. The operator may read and write values from some region of the graph around the active node, called the \textit{neighborhood} for that computation. An active node becomes inactive after the operator completes, although it may be reactivated later in the algorithm.

The global view of an algorithm is called the \textit{schedule}, and it consists of a description of the {\em locations of active nodes} in the graph and the {\em order} in which they must appear to have been processed by the implementation.

Algorithms generally locate active nodes in one of two ways. \textit{Topology-driven} algorithms operate in rounds; in each round, the operator is applied to all the nodes in the graph. Rounds terminate when some global convergence criterion is reached. The fast sweeping method is an example of a topology-driven algorithm. In contrast, \textit{data-driven} algorithms start with some initial set of active nodes. As active nodes are processed, other nodes may become active. The algorithm terminates when no active nodes remain to be processed. Implementations of data-driven algorithms use \textit{worklists} to keep track of active nodes. The fast marching method and fast iterative method are both examples of data-driven algorithms; the band tracked by each of those algorithms consists of the active nodes.

Ordering constraints on the processing of active nodes fall into two categories: \textit{unordered} and \textit{ordered}. In unordered algorithms, any order of processing active nodes will produce correct results. Although unordered algorithms allow processing active nodes in any order, they may use soft priorities to select active nodes for processing. These priorities are used only for efficiency since an unordered algorithm must produce correct results regardless of the order in which nodes are processed. In ordered algorithms, active nodes are required to appear to have been processed in a specific order. The fast marching method is an example of an ordered algorithm since only the earliest priority node is allowed to be processed. The fast iterative and fast sweeping methods also impose ordering constraints of some kind. In this paper, we argue that the deterministic properties of the Eikonal differencing scheme make it amenable to unordered scheduling with soft priorities.

\subsection*{TAO analysis of existing Eikonal solvers}

\begin{algorithm}
\caption{Eikonal Update Operator}
\label{alg:eikonal-operator}
\begin{algorithmic}[1]
\Function{UpdateNode}{$n$}
  \State $\arrivaltime_H \vcentcolon=$ smallest $\arrivaltime$ value at a horizontal neighbor of $n$
  \State $\arrivaltime_V \vcentcolon=$ smallest $\arrivaltime$ value at a vertical neighbor of $n$
  \If{$\abs{\arrivaltime_H - \arrivaltime_V} \geq \frac{h}{\speedfunction\pr{n}}$}
    \State $\arrivaltime\pr{n} \vcentcolon= \operatorname{min}\pr{\arrivaltime_H, \arrivaltime_V} + \frac{h}{\speedfunction\pr{n}}$
  \Else
    \State compute $\arrivaltime\pr{n}$ using Equation~\ref{eq:difference_as_root}
  \EndIf
\EndFunction
\end{algorithmic}
\end{algorithm}

In solvers for the Eikonal equation, the operator applies the differencing scheme to update an estimated solution value at a given node. This operation reads the values at the node's neighbors and then writes the value at that node, as shown in Algorithm~\ref{alg:eikonal-operator}. With respect to the terminology discussed in \cite{the_tao}, this is called a \textit{pull-style} vertex operator. The different Eikonal solvers use different schedules to apply this operator to nodes in the graph.

\subsubsection*{Fast Marching Method}
The fast marching method is a data-driven, ordered algorithm that closely resembles Djikstra's algorithm for the single source shortest path problem. It schedules its updates using a priority queue.
Tsitsiklis introduced the original fast marching method in \cite{fmm_original}.
The more commonly used variant based on upwind differencing was introduced by Sethian in \cite{fmm_main}, and a similar method was also introduced by Helmsen et. al. in \cite{fmm_also_discovered}.
The fast marching method was subsequently extended to triangulated domains by Sethian and others in \cite{fmm_triangulations} and \cite{fmm_triangulations_2}.
A second order variant was introduced in \cite{fmm_second_order} and a second order variant for unstructured meshes was introduced in \cite{fmm_second_order_unstructured}.
In \cite{fmm_variational}, Fomel provided a variational formulation of the fast marching method and demonstrated an equivalence between the Eikonal equation and a local version of Fermat's principle.
For simplicity, we will focus primarily on the first order upwinding scheme on regular grids introduced by Sethian \cite{fmm_main}; however the parallelization techniques described here are much more broadly applicable and translate naturally to the irregular mesh and higher order cases.

The fast marching method uses a priority queue to schedule updates to the estimated arrival time of nodes.
A node is put on the priority queue when its arrival time is updated, using the arrival time as the priority.
When a node is taken off of the priority queue, its neighbors are updated using the pull-style operator shown in Algorithm~\ref{alg:eikonal-operator}. Pseudocode for the fast marching method is shown in Algorithm~\ref{alg:fmm}.

\begin{algorithm}
\caption{Fast Marching Method}
\label{alg:fmm}
\begin{algorithmic}[1]
\State Initialize $\arrivaltime\pr{n} \vcentcolon= \infty$ for all grid points $n$
\State Initialize $\arrivaltime\pr{n}$ appropariately for boundary nodes
\State Push the boundary nodes onto the priority queue $w$
\While {$w$ is not empty}
  \State $n \vcentcolon=$ pop the smallest element of $w$
  \For {each neighbor $m$ of $n$}
    \State $\operatorname{UpdateNode}\pr{m}$
    \If{$\arrivaltime\pr{m}$ decreased}
      \State Push $m$ onto $w$ with priority $\arrivaltime\pr{m}$
    \EndIf
  \EndFor
\EndWhile
\end{algorithmic}
\end{algorithm}

In the literature on the fast marching method, the active nodes in the priority queue are often referred to as a {\em narrow band} that is said to ``march'' from the boundary through the domain.
These descriptions often also introduce tags to mark the state of points with respect to this narrow band, namely ``accepted", ``band", and ``far" \cite{sethian_fmm_survey}.
It is not necessary to explicitly track these states.
Implicitly, nodes with an arrival time that is smaller than the smallest entry in the queue are ``accepted".
Nodes that have a marked arrival time other than $\infty$ that are not final are ``band".
Nodes with an arrival time of $\infty$ are ``far".

The fast marching method has been described as ``inherently sequential'' \cite{fmm_parallel_7} since a node is allowed to update its neighbors only after all nodes with earlier arrival times have been processed. However, the fast marching method is known to be robust to some priority inversion.
For example \cite{fmm_sloppy} investigated using an ``untidy priority queue" and found minimal degradation in the computed solution when some priority inversions were allowed. Later in this paper we will discuss how allowing some priority inversions provides abundant opportunities for parallelism, why there may be observed changes in the computed solution when priority inversions are present, and how these errors can be mitigated or eliminated entirely.

\subsubsection*{Fast Iterative Method}
The fast iterative method \cite{fast_iterative} is a data-driven algorithm that proceeds in bulk-synchronous \cite{bsp_original} rounds.
The fast iterative method has shown good parallel performance in practice \cite{fmm_parallel_survey} and has even been extended to GPUs \cite{fim_meshes} \cite{fim_meshes_3d}.
There are no ordering constraints between nodes processed in the same round, but each round must complete before the next begins. This is
implemented by having two worklists; in each round, work items are processed from one worklist and newly created workitems are pushed on the other worklist (in the pseudocode, these worklists are called $w$ and $w'$ respectively). At the start of the algorithm, the current worklist $w$ is populated with the boundary nodes. When processing each item on the work list, the node is updated using the first-order differencing scheme, then, if its value has not decreased significantly (with respect to a user-specified threshold), its neighbors are also updated and pushed onto the work list if their values decreased. If its value has decreased significantly, the active node is pushed onto the work list for the next round. This approach is shown in Algorithm~\ref{alg:fim}

\begin{algorithm}
\caption{Fast Iterative Method}
\label{alg:fim}
\begin{algorithmic}[1]
\State Initialize $\arrivaltime\pr{n} \vcentcolon= \infty$ for all grid points $n$
\State Initialize $\arrivaltime\pr{n}$ appropariately for boundary nodes
\State Push the boundary nodes onto the work list $w$ for the first round
\While {$w$ is not empty}
  \State $w' \vcentcolon=$ empty work list
  \While {$w$ is not empty}
    \State $n \vcentcolon=$ pop an element from $w$
    \State $\operatorname{UpdateNode}\pr{n}$
    \If{$\arrivaltime\pr{n}$ decreased by less than a threshold $\epsilon$}
      \For {each neighbor $m$ of $n$}
        \If{$m$ in $w$}
          \Continue
        \EndIf
        \State $\operatorname{UpdateNode}\pr{m}$
        \If{$\arrivaltime\pr{m}$ decreased}
          \State Push $m$ onto $w'$
        \EndIf
      \EndFor
    \Else
      \State Push $n$ onto $w'$
    \EndIf
  \EndWhile
  \State $w \vcentcolon= w'$
\EndWhile
\end{algorithmic}
\end{algorithm}

Though the fast iterative method initially appears complex, it is simply an alternative schedule for the same local update.
It prioritizes updating nodes until their estimated arrival time stabilizes.
It will produce a single deterministic result since all active nodes will eventually be processed.
The existing work on the fast iterative method only shows consistency with the Eikonal equation when $\epsilon = 0$.
A nonzero $\epsilon$ value may cause some updates to not be propagated, but it is not currently clear what relation the computed result will have to the desired approximate Eikonal solution.

\subsubsection*{Fast Sweeping Method}
The fast sweeping method is a topology-driven ordered algorithm.
In \cite{finite_diff}, when this upwind differencing scheme was first introduced, the proposed update schedule is to iterate over all the nodes and iterate until convergence with respect to some tolerance.
This approach, shown in Algorthm~\ref{alg:eikonal-topological} is an example of a topology-driven schedule.
This approach trades some redundant work for not having to explicitly track active nodes.

\begin{algorithm}
\caption{Topology-driven Eikonal Solver}
\label{alg:eikonal-topological}
\begin{algorithmic}[1]
\State Initialize $\arrivaltime\pr{n} \vcentcolon= \infty$ for all grid points $n$
\State Initialize $\arrivaltime\pr{n}$ appropariately for boundary nodes
\Repeat
  \For {each node $n$}
    \State $\operatorname{UpdateNode}\pr{n}$
  \EndFor
\Until{no nodes were updated in the last iteration.}
\end{algorithmic}
\end{algorithm}

Problem-specific knowledge can provide an iteration order to help make each pull-style iteration significantly more effective (\cite{martin_barnes_hut} is an example of this).
That is the core idea behind the fast sweeping method \cite{fast_sweeping}.
The fast sweeping method relies on the fact that, for some problems, it is unusual for the characteristic curves of the result to change direction frequently.
In other words, the ordering that the fast marching method would have used to compute the finalized value for each node iterates over the domain of interest in more or less the same direction.
To take advantage of this, the fast sweeping method iterates over the domain of interest in alternating directions in the hope that one of the orderings used will be able to update the values at many nodes in a single pass.
Algorithm~\ref{alg:fsm} shows this approach.

\begin{algorithm}[ht]
\caption{Fast Sweeping Method (2D $M \times N$ grids)}
\label{alg:fsm}
\begin{algorithmic}[1]
\State Initialize $\arrivaltime\pr{n} \vcentcolon= \infty$ for all grid points $n$
\State Initialize $\arrivaltime\pr{n}$ appropariately for boundary nodes
\State Index grid points with (i, j)
\Repeat
  \State $\operatorname{UpdateNode}\pr{n_{i,j}}$ {\bf for} (i, j) {\bf in} (1:M, 1:N)
  \State $\operatorname{UpdateNode}\pr{n_{i,j}}$ {\bf for} (i, j) {\bf in} (M:1, 1:N)
  \State $\operatorname{UpdateNode}\pr{n_{i,j}}$ {\bf for} (i, j) {\bf in} (M:1, N:1)
  \State $\operatorname{UpdateNode}\pr{n_{i,j}}$ {\bf for} (i, j) {\bf in} (1:M, N:1)
\Until{No nodes were updated in the last iteration}
\end{algorithmic}
\end{algorithm}

The conventional wisdom has been that, while the FMM is ideal for worst-case work efficiency when handling poorly-behaved inputs, the FSM is faster in cases where the domain of interest is relatively simple and the computed solution shows relatively few changes in the direction of propagation \cite{marching_sweeping_multilevel} \cite{fast_sweeping_comparison}.
Some improvements that allow the fast sweeping method to skip some redundant updates are possible \cite{fsm_improved}.
The simpler scheduling required for the FSM has led to several parallel implementations \cite{fast_sweeping_parallel_1} \cite{fast_sweeping_parallel_2}.
In \cite{marching_sweeping_multilevel} Chacon and Vladimirsky introduced a hybrid scheme where the FMM was used for coarse-grained scheduling and the FSM was used for fine-grained scheduling, but the parallelism available using that method was not considered in detail.

%% file: schedule.tex
\section*{A Novel Parallel Marching Method}
\label{sec:schedule}

The previous section presented a unified view of FMM, FSM, and FIM by showing that they use the same
operator but different schedules.
Later in this paper, we show that the known causality and monotonicity properties of the differencing scheme used to implement Eikonal solvers enable the use of fully unordered scheduling.
Because of this, we are able to take advantage of recent advances in concurrent priority scheduling techniques that were originally developed for the single source shortest path problem and other graph applications.
These approaches avoid contention in a parallel setting by allowing some priority inversion instead of letting threads contend for a single earliest priority work item.

\subsection*{The Asynchronous Marching Method}
The notion of a concurrent priority scheduler was first introduced in \cite{obim}.
The scheduler introduced there is called \textit{Ordered By Integer Metric} (OBIM).
It provides low-contention priority-aware scheduling for unordered algorithms and handles issues like load balancing and grouping work items together into chunks to further reduce the cost of work list traffic. An OBIM worklist makes a best-effort
attempt to return an (approximately) earliest priority work-item but it treats priorities as \textit{soft} priorities,
permitting work-items to be processed even if they do not necessarily have the earliest priority
in the system. Intuitively, OBIM trades off some amount of redundant work for increased parallelism.

The OBIM implementation we used requires priorities to be integers and performs well only 
when there are many work items corresponding to each integer priority.
Since arrival times are floating point numbers, these numbers must be mapped into bins with corresponding integer priorities. Though more complicated techniques are possible, we utilized a simple scaling and rounding approach to obtain integer priority values from the arrival times. A version of the fast marching method adapted to use a concurrent priority scheduler is shown in Algorithm~\ref{alg:amm}. The worklist $w$ is an OBIM worklist in which arrival time is used as a \textit{soft} priority, unlike in
Algorithm~\ref{alg:fmm}, in which arrival time is used as a \textit{hard} priority. 

\begin{algorithm}
\caption{Asynchronous Marching Method}
\label{alg:amm}
\begin{algorithmic}[1]
\State Initialize $\arrivaltime\pr{n} \vcentcolon= \infty$ for all grid points $n$
\State Initialize $\arrivaltime\pr{n}$ appropriately for boundary nodes
\State Push the boundary nodes onto the OBIM worklist $w$
\While {$w$ is not empty}
  \State $n :=$ pop a work item from $w$
  \For {each neighbor $m$ of $n$}
    \State $\operatorname{UpdateNode}\pr{m}$
    \If{$\arrivaltime\pr{m}$ decreased}
      \State Push $m$ onto $w$ with priority $\floor{\arrivaltime\pr{m} / scale}$
    \EndIf
  \EndFor
\EndWhile
\end{algorithmic}
\end{algorithm}

The use of soft priorities in this algorithm relies on the fact that the first-order Eikonal differencing scheme is amenable to fully unordered scheduling. Indeed a careful implementation of this algorithm yields deterministic results in spite of significant nondeterminism present in the algorithm's intermediate states. 

%% file: precision.tex
\section*{Floating Point Precision Issues}
\label{sec:precision}

This section provides a proof that active nodes can be processed in any order to compute an approximate solution to the Eikonal equation and discusses issues that arise due to loss of precision with floating point arithmetic.
These numerical precision issues can be mitigated or entirely elminiated.
If they are not eliminated, they can give rise to observed differences in the solutions computed by different update schedules, as has been observed in some of the prior work in this area \cite{fmm_sloppy}.
They can also give rise to small nondeterministic errors in the computed output when fully asynchronous parallel scheduling is used.

\subsection*{Deterministic Properties}

The deterministic properties of the local update scheme make it so that an Eikonal solver computes the same solution regardless of the order in which updates are applied.
The local update scheme is known to be both \textit{monotone} and \textit{causal}.
Monotonicity in this case means that the differencing scheme will never compute an increase in the arrival time at a node in response to a decrease in the value of another node.
Causality in this case means that the value at each node depends only on other nodes with earlier arrival times.
We will prove here that causality and monotonicity together make it so that updates can be scheduled in any order while preserving the deterministic properties.
These properties also allow the fast marching method to achieve $\mathcal{O}\pr{n \log\pr{n}}$ worst case complexity.
Precision issues involving monotonicity in the presence of round-off error will be addressed later in this paper.

Causality has been described as each node depending only on other nodes with earlier arrival times than its own \cite{marching_sweeping_multilevel}.
Alternatively, in a causal update, information is said to flow from lower valued nodes to higher valued nodes \cite{sethian_fmm_survey} \cite{sethian_book}.
Causality can also be defined in terms of characteristic curves \cite{sethian_hamilton_jacobi}, which is useful in demonstrating how causality in the discrete setting arises from the properties of the underlying PDE.

Intuitively, the idea of causality is that the the arrival time at a given point in the domain can be computed using only the arrival times at earlier nodes.
Unfortunately, during execution of an Eikonal solver, it is not always clear which nodes will be earlier.
Fortunately causality still applies to the local updates that happen during execution of an Eikonal solver in the sense that each update uses only the arrival times that are earlier than its output. Formlly, causality can be defined as follows.
In addition to allowing fully asynchronous execution to produce deterministic results, monotonicity and causality also allow data-driven Eikonal solvers to mark a node as active only when one of its neighbors decreases to a value below the value currently stored on that node.

Let $\tau$ be an ordered set with a maximum element and $N$ be a finite set.
Let $F$ be the set of functions $N {\to} \tau$.
In the case of the Eikonal solvers discussed in this paper, $N$ is the set of interior grid points, $\tau$ is the set of possible arrival times i.e. $\br{0, \infty}$, and $F$ is the set of possible arrival time configurations for the whole set of grid points.

A {\em local update operator} is a function $D {:} N {\times} F {\to} F$ such that $D(n,f)(m) = f(m)$ for all $m \neq n$.

Given $f,g{:}N {\to} \tau$ and a local update operator $D$, let $U_{n,f} {=} \{u{\in}N {-} \set{n} | f(u) {<} D(n,f)(n)\}$. $U_{n,f}$ is sometimes called the {\em upwind set} of $n$ with respect to $f$.
$D$ is said to be {\em causal} if $U_{n, f}{=} U_{n,g}$ and $f(u){=}g(u)$ for all $u{\in}U_{n,f}$ implies $D(n,f) {=} D(n,g)$.

In the context of the Eikonal solvers discussed in this paper, $D$ is the operation of applying the Eikonal differencing scheme (from Equation~\ref{eq:finite_difference}) at a given interior point in $N$ to compute some new state in $F$.
This update operator is causal because the value at each node is computed using only $\arrivaltime_H$ (the horizontal neighbor with smallest known arrival time) and $\arrivaltime_V$ (the vertical neighbor with the smallest known arrival time).
The computed arrival time at a given node will only change if $\arrivaltime_V$ or $\arrivaltime_H$ changes, so it is invariant with respect to changes in nodes with values greater than or equal to $\operatorname{max}\pr{\arrivaltime_V, \arrivaltime_H}$.
When both $\arrivaltime_H$ and $\arrivaltime_V$ are used, the differencing scheme computes a value larger than both $\arrivaltime_H$ and $\arrivaltime_V$, so this implies causality.
In the case where $\arrivaltime_H$ and $\arrivaltime_V$ differ enough that a one-dimensional update is used, the computed value is invariant with respect to changes in nodes with values greater than or equal to whichever neighbor is used, implying causality in that case as well.
This update operator is also monotone.
An equivalent of monotonicity was proved for the differencing scheme discussed here in \cite{finite_diff}.
Monotonicity and causality are standard conditions on the local update operator when fast marching methods are used, so the order invariance described here also applies to a variety of existing irregular and higher-order methods.

\begin{theorem}
\label{thm:deterministic}
Let $f_0 \in F$ be the state that maps each $N$ to the maximal element of $\tau$.
For a causal and monotone update $D$, there exists a unique $f' \in F$ such that:
\begin{itemize}
 \item $f'$ is fixed under updates at any node, i.e. $D\pr{n, f'} = f'$ for any $n \in N$.
 \item All maximal sequences $f_0 \dots f_k$ such that $f_i = D\pr{n_i, f_{i - 1}}$ for some $n_i$ and $f_{i - 1} \neq f_i$ are finite in length and terminate in $f'$.
\end{itemize}
\end{theorem}
\begin{proof}

We proceed by showing that the set $\Delta$ of states reachable from $f_0$ through a finite number of applications of $D$ forms a finite partially ordered set with a maximum element.

We may define a partial order on $\Delta$ by saying that for states $f, g \in \Delta$, $f \leq g$ if for all $n \in N$, $f\pr{n} \leq g\pr{n}$.
Note that since $f_0$ maps every node to the maximal value of $\tau$, monotonicity guarantees that any update will result in a new state with a decreased value at some point.
This means any update is strictly decreasing with respect to the partial order on $\Delta$.

Now suppose by way of contradiction that $\Delta$ is infinite.
This means that there is some infinite sequence $m_i = m_1, \dots$ of nodes such that if $f_i = D\pr{m_i, f_{i - 1}}$, $f_i$ is an infinite sequence of distinct states in $\tau$.
Let $M \subset N$ be the set of points that appear infinitely often in the sequence $m_i$.
After sufficiently many steps, every update will happen only at points in $M$, so let $\bar{f}$ be some state such that all subsequent updates happen only at nodes in $M$ and such that an update at each point in $M$ has happened at least once since the last update to any point not in $M$.
Let $x$ be the point where $\bar{f}$ attains its minimum on $M$.
Causality implies that subsequent updates at $x$ can only change the value at $x$ if some other node has been updated to a value lower than $\bar{f}\pr{x}$.
Let $\tilde{f}$ be the first state where some node $y$ is updated to a value less than $\bar{f}\pr{x}$.
Causality of $D$ implies that $D\pr{y, \tilde{f}}\pr{y}$ is the same as would have been computed when an update at $y$ last occurred, but that is impossible since this must be a nontrivial update.
This means that a fixed point $f'$ must be reached after some finite number of updates.
Since $N$ is finite, this also implies that $\Delta$ is finite.

We now show $f'$ is unique.
Suppose, for contradiction, that there are two distinct states $f'$ and $f''$ in $\Delta$ that are invariant under updates at any node.
Let $n \in N$ be the point with smallest $f'\pr{n}$ where $f'$ and $f''$ disagree.
Without loss of generality, say that $f'\pr{n}$ is also smaller than any value of $f''$ where $f'$ and $f''$ disagree.
Since $D$ is causal, $D\pr{n, f'}\pr{n} = D\pr{n, f''}\pr{n}$ and thus $f'\pr{n} = f''\pr{n}$, which is a contradiction.

\end{proof}

Theorem~\ref{thm:deterministic} implies that any process consisting of a series of causal updates on a finite set will terminate, regardless of the order in which these updates are processed.
The only condition is that nontrivial updates continue to happen until convergence is reached.
This applies to all of the Eikonal solvers discussed in this paper, provided that they are run to completion.
There is one caveat, however.
Although this differencing scheme is monotonic in theory, when it is implemented with finite precision arithmetic it may no longer be truly monotonic.
This can result in small nondeterministic error in the computed results and can cause different implementations to produce slightly different results.
These errors can be mitigated or entirely eliminated using techniques discussed later in this paper.

\begin{theorem}
\label{thm:bound}
A fixed point described in Theorem~\ref{thm:deterministic} in at most $2^{\abs{N}} - 1$ steps.
\end{theorem}
\begin{proof}
We proceed by induction.
The case where $\abs{N} = 1$ clearly reaches the fixed point $f'$ after at most $1$ update.
Now say that a fixed point is reached after at most $2^{\abs{N} - 1} - 1$ updates for a set of points with size $\abs{N} - 1$.
Let $x$ be a point where $f'$ assumes its minimum value.
The $\abs{N} - 1$ case applies to the set of points $N - \set{x}$ with the update rule $D$ and $x$ set to its initial value, so at most $2^{\abs{N} - 1} - 1$ nontrivial updates can happen without running an update at $x$.
This also applies to the set of points $N - \set{x}$ with the update rule $D$ and $x$ set to its final value, so at most $2^{\abs{N} - 1} - 1$ nontrivial updates can happen after updating $x$.
Causality implies $x$ reaches its final value as soon as it is updated, this means that at most $2^{\abs{N}} - 1$ nontrivial updates can occur before reaching $f'$.
\end{proof}

Although we provide no proof here, another important consequence of monotonicity and causality is that the fast marching method finishes in at most $\mathcal{O}\pr{\abs{N} \log\pr{\abs{N}}}$ time.
Similarly, the topologically driven algorithm presented in \cite{finite_diff} where all nodes are processed in rounds is guaranteed to terminate in $\mathcal{O}\pr{\abs{N}^2}$ time.
Theorem~\ref{thm:bound} shows that, although the order of updates does not affect the end result of the algorithm, it may still have a strong effect on an algorithm's computational efficiency.
However, since efficiency is the goal, in a parallel setting there is room for a parallel scheduling system to allow some small number of priority inversions in exchange for a reduction in contention between threads of execution.
How effective this trade-off will be depends heavily on the hardware used and on the characteristics of the input data.
This opens an opportunity for further empirical work in improving concurrent priority scheduling techniques and verifying which approaches work best for different combinations of data and hardware.

\subsection*{Loss of Monotonicity With Floating Point Arithmetic}

Thus far we have shown that the Eikonal differencing operator can be applied to any order to reach a single deterministic result.
Previous papers, however, have noted slight differences in the output arising from different schedules (\cite{fmm_parallel_7}, \cite{fmm_parallel_6}, and \cite{fmm_sloppy} are all examples).
Here we show that these differences arise because of floating point precision loss in the differencing operator.
Finite precision computations in the differencing operator in some cases can make the actual computations non-monotonic.
The slight variations from truly monotonic behavior also cause small nondeterministic variation in the output of our asynchronous marching algorithm.
These precision issues can be mitigated by some reordering of the arithmetic operations, further mitigated by more careful tracking of work items, and fully eliminated with modified rounding semantics in the differencing operator.

A well-known way \cite{marching_sweeping_multilevel} to implement the differencing scheme in Equation~\ref{eq:finite_difference} is to compute the updated arrival time $\arrivaltime\ofpoint[i,j]{n}$ by letting
\begin{equation*}
\begin{aligned}
\arrivaltime_H &= \min\set{\arrivaltime\ofpoint[i-1,j]{n}, \arrivaltime\ofpoint[i+1,j]{n}}\\
\arrivaltime_V &= \max\set{\arrivaltime\ofpoint[i,j-1]{n}, \arrivaltime\ofpoint[i,j+1]{n}}
\end{aligned}
\end{equation*}
(as was done earlier in this paper) so that the update to $\arrivaltime\ofpoint[i,j]{n}$ can be computed as
\begin{equation*}
\frac{1}{2} \pr{\arrivaltime_H + \arrivaltime_V + \sqrt{\pr{\arrivaltime_H + \arrivaltime_V}^2 - 2 \pr{\arrivaltime_H^2 + \arrivaltime_V^2 - \pr{\frac{h}{\speedfunction\ofpoint[i,j]{n}}}^2}}}
\end{equation*}
For brevity, refer to this quantity as $d\pr{\speedfunction\ofpoint[i,j]{n}, h, \arrivaltime_H, \arrivaltime_V}$ and the corresponding finite precision equivalent computed using this formula as $\tilde{d}$.

\begin{table}[ht!]
\centering
\caption{$\tilde{d}$ fails to be monotonic when $\speedfunction\ofpoint[i,j]{n} = 0.5860617808911898$, $h = 1$, $\arrivaltime_H = 2949.952952954425$}
\begin{tabular}{ccc}
Case & $\arrivaltime_V$ & $\tilde{d}\pr{\speedfunction\ofpoint[i,j]{n}, h, \arrivaltime_H, \arrivaltime_V}$ \\
\midrule
A & 2951.6464609\textbf{466993} & 2951.6592100736580 \\
B & 2951.6464609071786 & 2951.65921007\textbf{44856} \\
\bottomrule
\end{tabular}
\label{tab:prec_1}
\end{table}

Although \cite{finite_diff} includes a proof that this differencing scheme is monotonic, this ceases to be true when it is implemented in finite precision IEEE 754 arithmetic \cite{ieee754}.
Table~\ref{tab:prec_1} gives one concrete example of when this formula fails to be monotonic.
In this case, although $\arrivaltime_V' < \arrivaltime_V$, it is also true that
$\tilde{d}\pr{\speedfunction\ofpoint[i,j]{n}, h, \arrivaltime_H, \arrivaltime_V} < \tilde{d}\pr{\speedfunction\ofpoint[i,j]{n}, h, \arrivaltime_H, \arrivaltime_V'}$

This nonmonotonicity arises within the square root.
Though the two terms shown within the square root are themselves monotonic in $\arrivaltime_V$ even when implemented with finite precision arithmetic, their difference may not be.
This is exacerbated by catastrophic cancellation that occurs when $\arrivaltime_H$ and $\arrivaltime_V$ are relatively large and close together.
Given that this is a differencing scheme that can be used with successively finer and finer meshes, $\arrivaltime_H$ and $\arrivaltime_V$ may often be close together, and that is precisely the case when the quadratic update is used at all.

This nonmonotonicity from the finite precision arithmetic is why various previous papers show observed differences between the results computed by solvers that differ only in the schedule for the updates.
It can also cause existing Eikonal solvers to produce results that are not truly fixed under further updates.
As is shown in \cite{fmm_sloppy}, this monotonicity is a more serious problem when more priority inversions are present.
When asynchronous nondeterministic scheduling is used, it is critical to reduce these errors since they can lead to noticeable nondeterministic error in the output.

\subsection*{Mitigation Strategies}

Since catastrophic cancellation exacerbates the numerical errors in computing $\tilde{d}$, one fruitful area for improvement is algebraic simplification to reduce the affects of catastrophic cancellation.
Though we are not currently aware of any way to obtain a truly monotonic finite precision version of $d$ purely through reordering the arithmetic of the differencing scheme, the following reordered variant (call it $\hat{d}$) drastically reduces nonmonotonic behavior in practice.
\begin{equation*}
\frac{1}{2}\pr{\arrivaltime_H + \arrivaltime_V + \sqrt{2 \pr{\frac{h}{\speedfunction\ofpoint[i,j]{n}}}^2 - \pr{\arrivaltime_H - \arrivaltime_V}^2}}
\end{equation*}

\begin{table}[ht!]
\centering
\caption{$\hat{d}$ fails to be monotonic when $\speedfunction\ofpoint[i,j]{n} = 1500$, $h = 1.25$, $\arrivaltime_H = 0.05752086379104517$}
\begin{tabular}{ccc}
Case & $\arrivaltime_V$ & $\tilde{d}\pr{\speedfunction\ofpoint[i,j]{n}, h, \arrivaltime_H, \arrivaltime_V}$ \\
\midrule
A & 0.0579522029351838\textbf{1} & 0.05828490263459645 \\
B & 0.05795220293518380 & 0.0582849026345964\textbf{6} \\
\bottomrule
\end{tabular}
\label{tab:prec_2}
\end{table}

This version is still not perfectly monotonic since, in some cases a decrease in either $\arrivaltime_H$ or $\arrivaltime_V$ results in an increase in the square root term.
The conflicting increase and decrease of different terms, in the presence of finite precision arithmetic is still sufficient to cause nonmonotonic behavior in some cases.
Table~\ref{tab:prec_2} gives one concrete example of when this happens.
As was the case with the previous example, although $\arrivaltime_V' < \arrivaltime_V$ it is also true that
$\hat{d}\pr{\speedfunction\ofpoint[i,j]{n}, h, \arrivaltime_H, \arrivaltime_V} < \hat{d}\pr{\speedfunction\ofpoint[i,j]{n}, h, \arrivaltime_H, \arrivaltime_V'}$.
On the other hand, this simple mitigation, when used in conjunction with our concurrent priority scheduling technique, is sufficient to drastically reduce any error resulting from missed work items as is shown in Table~\ref{tab:error}.

It is possible to further mitigate precision loss through more careful tracking of which nodes become active.
Since the differencing scheme is known to be monotonic, it is common in many implementations to only check if the value at a point has decreased in order to decide whether or not to mark that point's neighbors for updates.
Instead checking if the value at a given point has changed does further reduce the error in the final result.
The theoretical guarantees with regards to termination no longer apply in this case, but in practice nonmonotonic behavior from the differencing scheme is so rare that this results in very few additional updates beyond those that would have been computed without this correction.
Table~\ref{tab:error} shows that the error remaining when using the rearranged differencing scheme can be further reduced through this more careful approach to tracking which points need to be updated.

\subsection*{Achieving Exact Monotonicity with Finite Precision Arithmetic}

Although there's no currently known way to achieve exact monotonicity via an explicit formula implemented in finite precision arithmetic, it is still possible to get a perfectly monotonic implementation of the differencing operator.
This can be done by redefining the roundoff semantics of the differencing scheme relative to a root finding problem that does behave in a monotonic way when implemented with finite precision arithmetic.
The approach described here applies to the first-order differencing scheme discussed in this paper, but it also provides a road map for how truly monotonic behavior might be obtained when working with a variant of the fast marching method that uses a more complicated numerical scheme.

To see how this is possible, note that the difference scheme used when $\abs{\arrivaltime_H - \arrivaltime_V} < \frac{h}{\speedfunction\ofpoint[i,j]{n}}$ can be rewritten as finding a value $\arrivaltime\ofpoint[i,j]{n}$ such that $\alpha\pr{\arrivaltime\ofpoint[i,j]{n}, \arrivaltime_H, \arrivaltime_V} = 0$ where
\begin{equation*}
\alpha\pr{\arrivaltime\ofpoint[i,j]{n}, \arrivaltime_H, \arrivaltime_V} = \pr{\arrivaltime\ofpoint[i,j]{n} - \arrivaltime_H}^2 + \pr{\arrivaltime\ofpoint[i,j]{n} - \arrivaltime_V}^2 - \frac{h^2}{\speedfunction\ofpoint[i,j]{n}}
\end{equation*}
Writing the differencing equation in this form has the beneficial property that, whenever $\arrivaltime\ofpoint[i,j]{n} >= \arrivaltime_H$ and $\arrivaltime\ofpoint[i,j]{n} >= \arrivaltime_V$, $\alpha$ responds monotonically to changes in $\arrivaltime\ofpoint[i,j]{n}$, $\arrivaltime_H$, and $\arrivaltime_V$.
In the case of $\arrivaltime_H$ and $\arrivaltime_V$ $\alpha$ decreases when either input increases.
Even when computing with finite-precision arithmetic, $\alpha$ either decreases or remains the same in response to an increase in either $\arrivaltime_H$ or $\arrivaltime_V$ as long as both values continue to be smaller than $\arrivaltime\ofpoint[i,j]{n}$.
This is because $\arrivaltime_H$ and $\arrivaltime_V$ each only appear once when computing $\alpha$ and squaring and addition with a constant are both monotonic operations even when implemented with IEE 754 floating point numbers.
In the case of $\arrivaltime\ofpoint[i,j]{n}$, both $\pr{\arrivaltime\ofpoint[i,j]{n} - \arrivaltime_H}^2$ and $\pr{\arrivaltime\ofpoint[i,j]{n} - \arrivaltime_V}^2$ do not decrease with an increase in $\arrivaltime\ofpoint[i,j]{n}$, so neither does their sum.

Now let $\bar{d}\pr{\arrivaltime_H, \arrivaltime_V, \speedfunction\ofpoint[i,j]{n}, h}$ be the difference scheme $d$ computed by selecting the smallest floating point value for $\arrivaltime\ofpoint[i,j]{n}$ with $\arrivaltime\ofpoint[i,j]{n} \geq \arrivaltime_H$ and $\arrivaltime\ofpoint[i,j]{n} \geq \arrivaltime_V$ such that $\alpha\pr{\arrivaltime\ofpoint[i,j]{n}, \arrivaltime_H, \arrivaltime_V, \speedfunction\ofpoint[i,j]{n}, h}$ is nonnegative.
$\bar{d}$ is monotonic even when finite precision arithmetic is used to compute $\alpha$.
Computing using these specific rounding semantics can be done with standard techniques.

Although it is possible to achieve perfect monotonicity, in-practice we have observed that Newton's method run on the root finding problem $\alpha = 0$ with a starting guess computed by the improved explicit formula is sufficient to obtain fully deterministic results.
This approach does not precisely follow the roundoff semantics described for computing $\bar{d}$, however it was sufficient to obtain deterministic results in the cases that we tested and a similar approach may prove to be good enough in other settings.
This approach incurs only a slight computational cost over using the explicit formula, as can be seen in Figure~\ref{fig:scaling_vs_sequential}.

%% file: results.tex
\section*{Empirical Results}
\label{sec:results}

We implemented our \obimfmm[abbr] algorithm in Galois, the state-of-the-art graph analytic system, and tested it with the Elastic Marmousi dataset, a standard test dataset in seismology (see Materials and Methods).

\begin{figure}[ht]
\centering
\includegraphics[width=\localfigwidth]{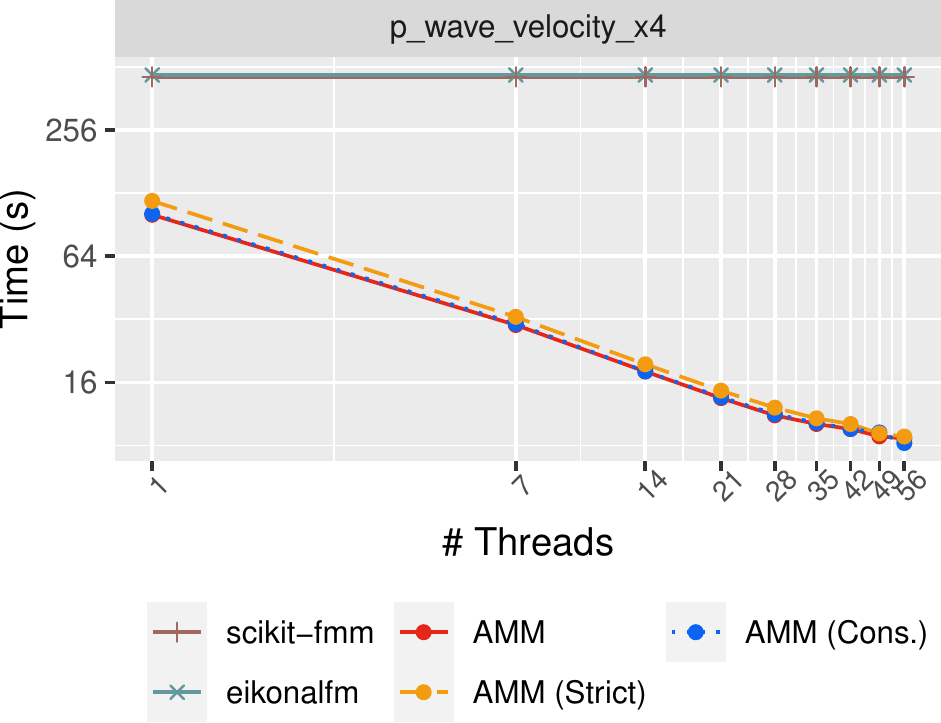}
\caption{Execution times of \obimfmm[abbr] compared with existing sequential FMM implementations.}
\label{fig:scaling_vs_sequential}
\end{figure}

Here we compare our implementation with two publicly available sequential first-order FMM implementations, \texttt{scikit-fmm}~\cite{scikit-fmm} and \texttt{eikonalfm}~\cite{eikonalfm}.
Figure \ref{fig:scaling_vs_sequential} shows the execution time with different numbers of threads.
The single-threaded execution of our implementation already beats the performance of the sequential baselines.
Our implementation achieves up to 54x speedup when using 56 threads.
The techniques for mitigating or fixing the nondeterministic floating point errors add relatively little cost, though there is some small overhead associated with obtaining fully deterministic results.

\begin{figure}[ht]
\centering
\includegraphics[width=\localfigwidth]{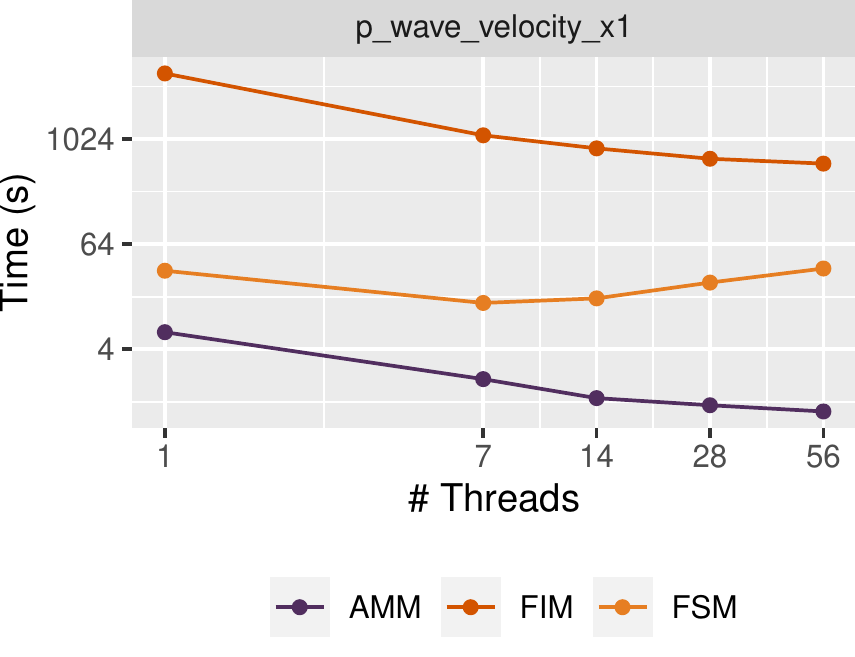}
\caption{Execution time of \obimfmm[abbr] compared with FSM and FIM (also implemented in Galois, see Materials and Methods).}
\label{fig:time_methods}
\end{figure}

Figure \ref{fig:time_methods} compares our OBIM-based FMM with FSM and FIM.
Our implementation shows not only good parallel scalability but also much better absolute performance.
Even in the single-threaded case it shows a noticeable improvement over the fast marching method since it does not pay the overhead of differentiating between work items with extremely similar priorities.
These results compare favorably with known shared-memory parallel Eikonal solvers \cite{fmm_parallel_survey}\cite{fmm_parallel_latest}.

\begin{figure}[ht]
\centering
\includegraphics[width=\localfigwidth]{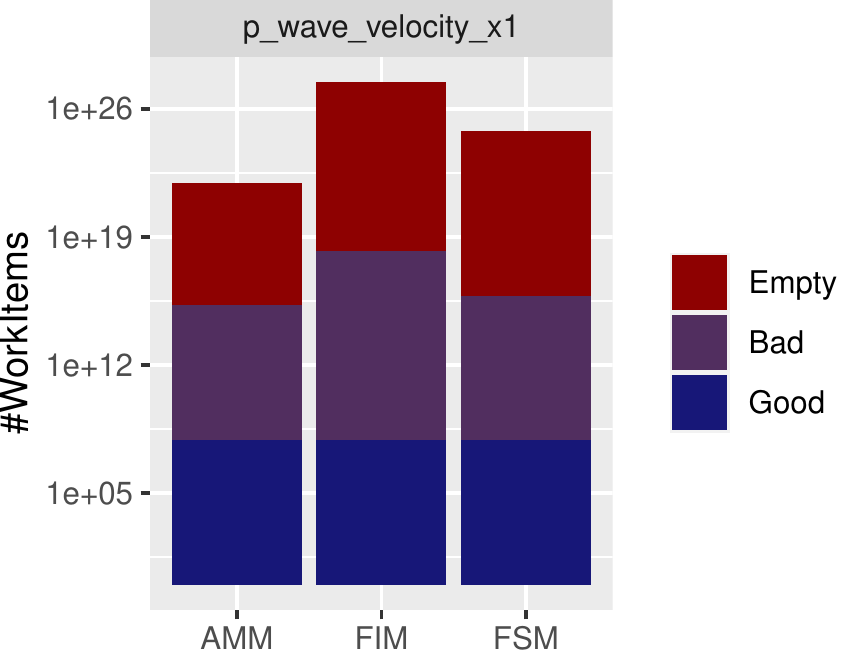}
\caption{Work items processed during the computation of different solvers.}
\label{fig:work_item}
\end{figure}

To compare the work efficiency of the methods we measured the number of good, empty, and bad updates processed by each algorithm.
This concept was originally applied to the SSSP problem in \cite{obim}.
A good update to a point is one that sets that point to its final value.
An empty update is when the differencing scheme is run at at point, but its value is not updated.
A bad update is when the value at a point is set to something other than that point's final value.
Figure \ref{fig:work_item} shows that, although AMM introduces more computational redundancy, there are many fewer bad updates since threads focus primarily on processing early-priority work items.
The \textit{bad} works have a greater impact on slowing down the execution than \textit{empty} works do, therefore the overhead in computation redundancy pays off by the parallelism gained from the OBIM scheduler.

\begin{table}[ht!]
\centering
\caption{Error with different mitigation strategies relative to the discrete set of constraints imposed by the differencing scheme.}
\begin{tabular}{ccc}
Mitigation Strategy & Input Scale & Error \\
\midrule
no mitigation & x1 & $10^{-12}$ \\
rearranged arithmetic & x1 & $10^{-16}$\\
rearranged arithmetic & x4 & $10^{-16}$\\
rearranged with conservative work tracking & x4 & $10^{-16}$\\
true monotonicity & x4 & $0$\\
\bottomrule
\end{tabular}
\label{tab:error}
\end{table}
Table~\ref{tab:error} shows the error relative to the discrete set of constraints on each node imposed by the differencing scheme.
The listed error is the maximum relative error between the final value at a point and the value given by applying the differencing scheme at that point again using the final values of its neighbors.
This is a measure of how well the computed solution satisfies the differencing scheme at each point.
It does not measure how accurate the differencing scheme may be for computing a solution to the underlying continuous shortest path problem.
Table~\ref{tab:error} shows that a true fixed point can be reached if a truly monotone update is used.
It also shows that arithmetic reordering in this case drastically reduces the observed error in most cases.
The conservative work item tracking mitigation was only a slight improvement over the reordered arithmetic{\textemdash}showing $0$ error when used with fewer threads, but jumping back up to the same error as before during some runs with more threads.

%% file: conclusion.tex
\section*{Conclusion}
\label{sec:conclusion}

In this paper we have provided a unified framework for existing first-order Eikonal equation solvers that describes the fast marching method \cite{fmm_main}, the fast iterative method \cite{fast_iterative}, and the fast sweeping method \cite{fast_sweeping} all as different scheduling techniques for the same local update.
We have shown that these local updates can be processed in any order and that, with careful handling of roundoff error in the differencing operator, different schedules can result in identical results.
We have focused our discussion of monotonicity on the first-order upwind differencing scheme for the Eikonal equation, however our treatment of the corresponding precision issues provides a roadmap for how to handle these concerns in the future and opens the way for future research to explore how monotonicity can be enforced for higher-order and irregular schemes.
We have investigated applying state-of-the-art concurrent priority scheduling techniques, originally developed for SSSP \cite{obim}, to Eikonal equation solvers and have demonstrated excellent performance in the shared-memory parallel setting on CPUs.
These results reinforce the idea that the key concern for achieving good performance in an Eikonal solver is to match the scheduling strategy to the expected hardware and input data.
This demonstrates a need in this field for a suite of standard test data and problems representative of the kinds of priority distributions that are expected to occur when running an Eikonal solver.
It also motivates further research into concurrent priority scheduling techniques and helps characterize potential scheduling approaches that could be useful on GPUs and other diverse hardware platforms. 

%% file: methods.tex
Our experiments were performed on a 56-core machine with 4 Intel(R) Xeon(R) Gold 5120 processors (14 cores each running at 2.20GHz).

The Elastic Marmousi dataset \cite{marmousi2} contains a speed map of $2801 \times 13601$ sample points with a spacing of 1.25 meters between samples. To better illustrate the parallel scaling of our implementation, we upsampled the Marmousi dataset by a factor of 4 on each axis using nearest neighbor interpolation to obtain a $11204 \times 54404$ grid of values.

FSM and FIM in Figure~\ref{fig:time_methods} are implemented in Galois as we did for our algorithm.

For FSM, \cite{fast_sweeping_parallel_1} parallelizes the four sweepings of different ordering with four CPUs and exploits the domain decomposition approach to promote the parallelism.
The proposed ``additive'' and ``multiplicative'' versions uses inter- and intra-subdomain parallelism respectively.
\cite{fast_sweeping_parallel_2} adopts the Cuthill-McKee ordering that gives independent sets of points for each sweeping, which potentially reduces iterations and the memory usage.
We follow the algorithm presented in \cite{fast_sweeping_parallel_2} for comparison.

For FIM, we follow the algorithm presented in \cite{fast_iterative} in general but avoid checking whether a work item is present in the worklist.
We allow repeated work items and push items into the worklist with their current solution value as a tag.
The fact that repetitive pushes happen when the solution value of the point gets updated implies that we can remove the repetition by checking if the tag is stale compared to the latest solution value of the point.
This implementation preserves the ``is-in-worklist'' semantics but eliminates the thread contention for maintaining the ``in-list'' state for each point.